\documentclass[aps,pra,twocolumn,superscriptaddress,showpacs,nofootinbib]{revtex4}
\usepackage{graphicx}
\usepackage{bm}
\usepackage{amssymb}
\usepackage{epstopdf}
\usepackage{amsmath}
\usepackage{amstext}
\usepackage{amsthm}
\usepackage{amsfonts}
\usepackage{latexsym}
\usepackage{hyperref}
\usepackage[margin,index]{fixme}
\usepackage[T1]{fontenc}

\newcommand{\ket}[1]{\left\vert#1\right\rangle}
\newcommand{\bra}[1]{\left\langle#1\right\vert}

\newcommand{\av}[2]{\left\langle#1\right\rangle_{#2}}

\newcommand{\ketbra}[2]{\ket{#1}\!\!\bra{#2}}

\newcommand{\dd}{\mathrm{d}}

\newcommand{\id}{\boldsymbol{1}}

\newcommand{\Tr}{{\rm tr}}
\renewcommand{\emph}[1]{{\it #1}}
\renewcommand{\vec}[1]{\boldsymbol{#1}}
\newcommand{\ver}[1]{\boldsymbol{\hat #1}}

\begin{document}

\title{Long-range multipartite entanglement close to a first order quantum phase
transition}
\author{J. Stasi\'nska}
\affiliation{Departament de F\'isica. Universitat Aut\`{o}noma de Barcelona, E08193
Bellaterra, Spain}
\affiliation{ICFO-Institut de Ci\`{e}ncies Fot\`{o}niques.  E08860 Castelldefels, Spain}
\author{B. Rogers}
\author{M. Paternostro}
\author{G. De Chiara}
\affiliation{Centre for Theoretical Atomic, Molecular and Optical Physics,
Queen University Belfast, Belfast BT7 1NN, United Kingdom}
\author{A. Sanpera}
\affiliation{ICREA-Instituci\`o Catalana de Recerca i Estudis Avan\c{c}ats, E08011
Barcelona} 
\affiliation{Departament de F\'isica. Universitat Aut\`{o}noma de Barcelona, E08193
Bellaterra, Spain}

\begin{abstract}
We provide a novel insight in the quantum correlations structure present in strongly correlated systems 
beyond the standard framework of bipartite entanglement. To this aim we first exploit rotationally
invariant states as a test bed to detect genuine tripartite entanglement beyond the nearest-neighbor in
spin-1/2 models. Then we construct in a closed analytical form a family of entanglement witnesses which provides a sufficient condition to determine if a state of a many-body system formed by an arbitrary number of spin-1/2 particles possesses genuine tripartite entanglement, independently of the details of the model. We illustrate our method by analyzing in detail the anisotropic XXZ spin chain close to its phase transitions, where we demonstrate the presence of long range multipartite entanglement near the crtical point and the breaking of the symmetries associated to the quantum phase transition.
\end{abstract}
\maketitle

\section{Introduction}\label{sec:intro}
The characterization of entanglement in many-body strongly correlated systems 
has been a very active research area in the last decade (see e.g. Ref.~\cite{Amico2008} for a review). 
Entanglement  is expected to be particularly relevant in quantum phase transitions (QPT) and, in order to gather valuable insight, ground states of paradigmatic spin chain models have been exhaustively analyzed.  First, it was shown in Refs.~\cite{Osborne2002,Osterloh2002} that for the Ising chain in a transverse field, pairwise entanglement measured by the concurrence between nearest (and next-to nearest) neighbors signals the position of the critical point (but does not display critical behavior) while is strictly zero otherwise. The conjecture that, near criticality, entanglement should be present at all length-scales led to the concept of localizable entanglement, which was introduced in Ref.~\cite{Verstraete2004c}. For a given $N$-partite state $\rho_{1\,2\dots\,N}$, the localizable entanglement is the maximum average entanglement that can be made available to two pre-determined parties (say, spins 1 and 2), by performing general local
measurements on the rest of the system. If the correlation length diverges at criticality, so does the length of localizable entanglement. However, the converse is not necessarily true~\cite{Verstraete2004c}. 

A quite different approach was considered in Ref.~\cite{blockentropy}, where the scaling of the entanglement between a given block of spins and the rest of a chain was analyzed against the size of the block itself. In such cases, the entanglement between two blocks can be fully determined simply by considering the von Neumann entropy of a block (also termed entanglement entropy). In this approach, contrary to the pairwise case, entanglement is clearly related to critical behavior. Away from criticality, the entanglement entropy reaches a constant value and shows logarithmic divergence with the size of the block when approaching the critical point. Conformal field theory was used to relate such divergence to the central charge of the corresponding effective theory, thus linking entanglement and the universality class of the corresponding QPT.  Remarkably, both the entanglement entropy as well as the Renyi entropies are functions of the eigenvalues of the reduced density matrix of the block.

While all previous studies concern exclusively bipartite entanglement (either pairwise or block-block) a full description of many-body strongly correlated systems should include multipartite entanglement. Indeed, multipartite entanglement has been demonstrated in certain spin models~\cite{Wang2002,Stelmachovic2004,Bruss2005,Guhne2005,Guhne2006} and its role in QPT has been long discussed (see e.g. Ref.~\cite{deOliveira2006} and references therein).  However, the study of multipartite entanglement is presently much less developed in light of its daunting nature and the lack of appropriate tools. In fact, we even lack general measures of entanglement for mixed states of three spins and it is not possible to extend the concept of entanglement entropy to more than two blocks.

In this paper we analyze long range multipartite entanglement in the vicinity of quantum phase transitions providing a general method that relies only on three-point correlators only. The paper is organized as follows: In Sec. \ref{sec:rotinv} we briefly review rotationally invariant states that will be used as the starting point for the construction of a family of entanglement witnesses (i.e., observables) in Sec. \ref{sec:witness}. We provide a sufficient condition to assess genuine tripartite entanglement in the ground state of a many-body system. By relying on the availability of three-point correlation functions only, our approach does not depend on the Hamiltonian properties (type and range of the interactions, symmetries of the model, lattice geometry), its dimensionality or the actual choice of the subset of three spins picked from the lattice. With such tools at hand, we demonstrate that genuine multipartite entanglement (GME) is highly sensitive to quantum phase transitions. In order to provide a significant context where our formalism can be applied, we focus on the spin-$1/2$ XXZ chain. In Sec. \ref{sec:XXZ} we first shortly review the properties of the model. Then we consider the transition from the XY critical phase to the ferromagnetic one. It is known that as the critical point is approached pairwise bipartite entanglement becomes independent of the distance between the spins~\cite{verrucchi}. Here we show that also long range multipartite entanglement emerges in the vicinity of the the critical point. In this sense we provide a finer-grained entanglement structure than previous analysis based on collective operators \cite{Bruss2005,Guhne2005} by proving that entanglement extends far beyond the nearest-neighbor scale. We summarize our results in Sec. \ref{sec:summary}.

\section{Rotationally invariant states}\label{sec:rotinv}
Our analysis stems from rotationally invariant states of three qubits, whose
entanglement properties have been unambiguously characterized in Ref.~\cite{Eggeling2001} through a set of scalar inequalities. 
By projecting a generic state onto its rotationally invariant subspace, we can
also address GME in non-rotationally invariant states. Equivalently, one can
detect GME by constructing suitable rotationally invariant entanglement witnesses. 

Before proceeding further we review, for completeness, the characterization of rotationally invariant
states. We then provide a geometrical description of such space and construct
a family of entanglement witnesses. 


The class of $SO(3)$ invariant tripartite states $\rho$ is defined as
\begin{equation}
\left\{\rho: \underset{\ver{n},\theta}{\forall}\;
\left[\mathcal{D}^{j_1}_{\ver{n},\theta}\otimes
\mathcal{D}^{j_2}_{\ver{n},\theta}\otimes
\mathcal{D}^{j_3}_{\ver{n},\theta},\rho\right]=0\right\},
\end{equation}
where $\mathcal{D}^{j_i}_{\ver{n},\theta}$ denotes the unitary irreducible
representation of the rotations $R(\ver{n},\theta)$ from the $SO(3)$ group.

For three qubits, $\rho$ acts trivially in the two subspaces of total angular momentum $1/2$ and the one of $3/2$. However, since in general $[\rho,\bm J_{12}]\neq 0$ (with  $\bm J_{12}=\bm j_{1}+\bm j_{2}$) the two subspaces with  $J=1/2$ can be mixed.  
Denoting by $P_{1/2,a(b)}(\theta, \phi) $ the two orthogonal projectors onto the mixed 
$J=1/2$ subspaces, parametrized by the angles $\theta$ and $\phi$, and by $P_{3/2}$ the
projector onto the subspace $J=3/2$, the density operator can be decomposed as
\begin{equation}
\label{eq:ourrep}
\rho=\frac p2 P_{1/2,a}(\theta, \phi) + \frac q2 P_{1/2,b}(\theta,
\phi)+\frac{1-p-q}{4}P_{3/2},
\end{equation}
where $0\le p,q\le 1$. Hence $4$ real parameters  $p,q,\theta,\phi$ describe the set of
three-qubit $SO(3)$-invariant states. The construction can be extended to higher spins, 
although  the number of parameters grows dramatically. For instance, for spin-$1$
particles 13 variables are required. As the $SO(3)$ and $SU(2)$ groups are isomorphic, the
above representation can be straightforwardly mapped onto the one for $SU(2)$ invariant
states~\cite{Eggeling2001}
\begin{equation}
\label{eq:wernerrep}
\rho=\frac14\sum_{k=+,0,1,2,3} r_k\;\; R_k,
\end{equation}
where $r_k=\Tr(\rho R_k)$ and the factor $1/4$ ensures normalization. The Hermitian operators $R_k$ read 
\begin{subequations}
\begin{align}
R_+&=\left(\id +V_{12}+V_{23}+V_{13}+V_{123}+V_{321}\right)/6,\\
R_0&= \left(2 \id-V_{123}-V_{321}\right)/3,\\
R_1&= \left(2 V_{23}-V_{13}-V_{12}\right)/3,\\
R_2&=\left(V_{12}-V_{13}\right)/{\sqrt 3},\\
R_3&=i\left(V_{123}-V_{321}\right)/\sqrt{3},
\end{align}
\end{subequations}
where $V_{ij}$ is the permutation (or swap) operator acting on qubits $i$ and $j$; $V_{123}=V_{12}V_{23}$ and $V_{321}=V_{23}V_{12}$
are the two operators which cyclically permute all three particles, and $\id$ denotes the identity. The operator $R_{+}$ ($R_0$) is proportional to the projector $P_{3/2}$ ($P_{1/2}=P_{1/2,a}+P_{1/2,b}$). The three remaining matrices $R_i$ ($i=1,2,3$) act on the four-dimensional subspace of total spin-$1/2$, follow the angular
momentum commutation rules and are thus traceless. In order to ensure that Eq.~\eqref{eq:wernerrep} represents a legitimate state, the coefficients $r_k$ must satisfy the conditions 
\begin{equation}
r_+,r_0 \geqslant 0,\; r_+ + r_0=1,\;
r_1^2+r_2^2+r_3^2\leqslant  r_0^2.
\end{equation}
\section{Multipartite entanglement characterization and detection}\label{sec:witness}
The entanglement characterization of three-qubit states distinguishes four classes
\cite{Dur2000,Acin2001,DeVicente2012}:
(i) separable states $\mathcal S$  of the form $\rho=\sum_i \lambda_i\,
\rho_{i}^{(1)}\otimes \rho_{i}^{(2)}\otimes\rho_{i}^{(3)}$, (ii) biseparable states
$\mathcal B$ belonging to the convex hull of states separable with respect to one of the
partitions $1|23$, $2|13$ or $3|12$ denoted by $\mathcal B_1$, $\mathcal B_2$, $\mathcal
B_3$ respectively, and two GME classes (iii) W-type states, and (iv) GHZ-type states. Each
class embraces those that are lower in the hierarchy, i.e. $\mathcal S\subset \mathcal B
\subset \rm{W} \subset \rm{GHZ}$. The distinction between the W-type states and the
GHZ-type ones arises from the fact that for, three qubits, there are two non-equivalent
classes of genuinely entangled states with representative elements being precisely the
W state $ \ket{\rm W}=1/\sqrt{3}(\ket{100}+\ket{010}+\ket{001})$, and the GHZ state
$\ket{\rm GHZ}=1\sqrt{2}(\ket{000}+\ket{111})$. Elements of one class cannot be
inter-converted into elements of the second one using stochastic local operations and
classical communications. Therefore, the W and GHZ classes are formed by convex
combinations of states equivalent to $\ket{\rm W}$ and of combinations of states
equivalent to $\ket{\rm GHZ}, respectively$.

In Ref.~\cite{Eggeling2001}, the subsets $\mathcal S$ and
$\mathcal B$ of the $SU(2)$ invariant states are fully described in terms of inequalities
for the coefficients $r_k$. In particular, the set $\mathcal S$ is constrained by the conditions 
\begin{eqnarray}
&1/4 \leqslant r_+ \leqslant 1,&\nonumber\\ 
&3 r_3^2 +(1-3r_+)^2 \leqslant (r_1+r_+)[(r_1-2r_+)^2-3r_2^2].& 
\end{eqnarray}
Analogously, the states belonging to the set $\mathcal
B_1$, can be shown to fulfill the condition 
\begin{eqnarray}
&|m|<1,&\nonumber\\
&3 (r_2^2+r_3^2) \leqslant (1-|m|)^2-[(r_1-r_+)-|m|]^2,&
\end{eqnarray}
where $m=1+r_1-2r_+$. The corresponding sets
$\mathcal B_2$ and $\mathcal B_3$ are found by rotating $\mathcal B_1$ by $\pm2\pi/3$
around the axis $r_0$. 
Finally, the set $\mathcal{T}$ of genuine tripartite entangled states is found
as the complement of $\mathcal B$ in the set of all states.

\begin{figure}[t]
\begin{center}
(a)\hspace{0.43\columnwidth} (b)\\
\includegraphics[width=0.23\textwidth]{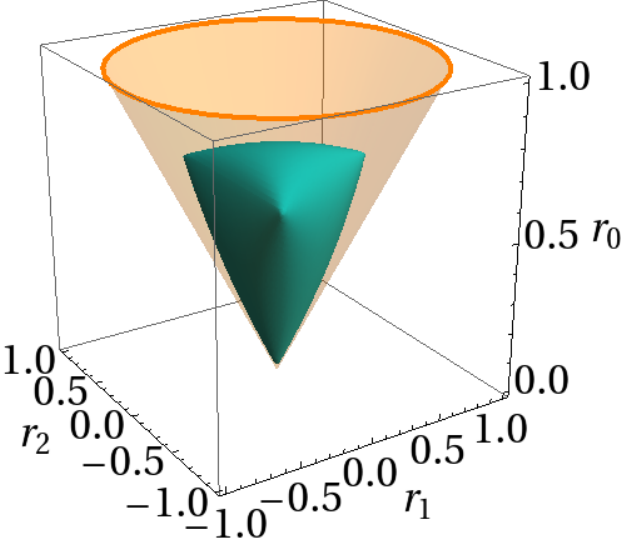}
\includegraphics[width=0.23\textwidth]{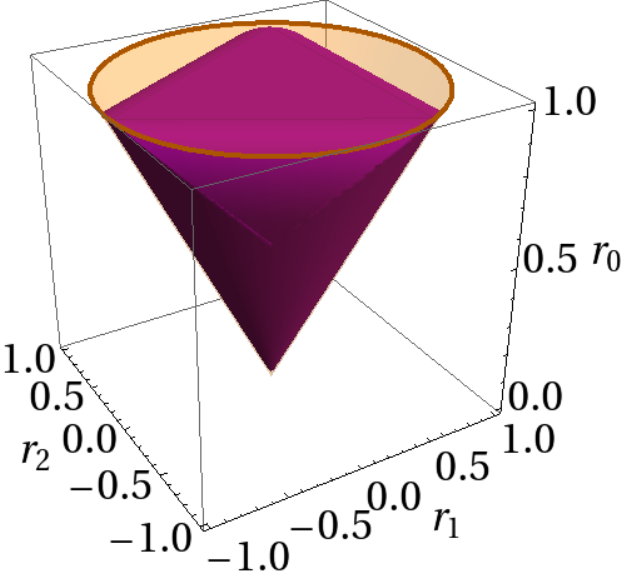}

(c)\hspace{0.43\columnwidth} (d)\\
\includegraphics[width=0.24\textwidth]{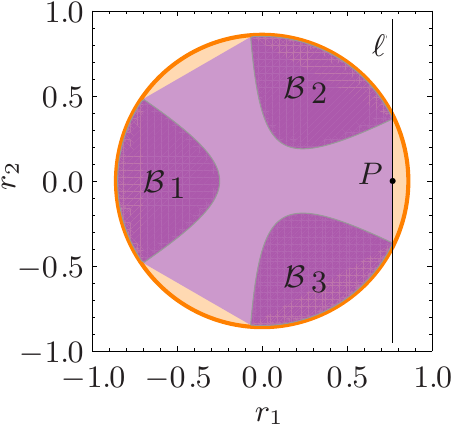}
\includegraphics[width=0.23\textwidth,trim=0 -1 0 0]{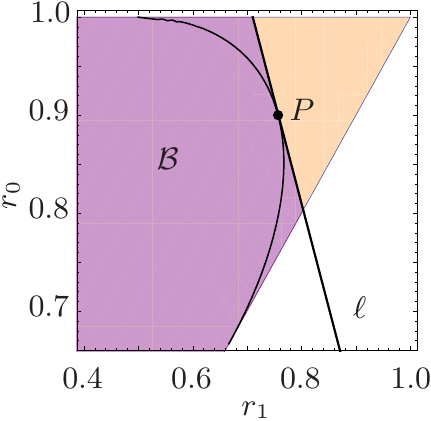}
\caption{(Color online) Graphical representation of the set of real rotationally
invariant states and its subsets with various types of entanglement in the space of dimensionless parameters $r_0,r_1,r_2$. (a) Separable and (b) biseparable states. (c) A horizontal cut showing the three sets of biseparable states, their convex hull and the tangent in the point $P$ to the set representing the witness. (d) Same as in (c) but a vertical is shown.}
\label{fig:sets}
\end{center}
\end{figure}

For real Hamiltonians, the above description is further simplified as their ground states and their reductions are represented by real
density operators. This is equivalent to setting $\phi=0$ in (\ref{eq:ourrep}) or $r_3=0$
in (\ref{eq:wernerrep}) and allows us to visualize the set of rotationally invariant
states in the space $r_1,r_2,r_0$ (Fig. \ref{fig:sets}). The complete set 
is a cone with symmetry axis parallel to the axis $r_0$. In  Fig. \ref{fig:sets}(a-b) we
depict the set of separable $\mathcal S$ and biseparable  $\mathcal B$ states,
respectively; the complementary volume contains genuine tripartite entangled states.
Fig. \ref{fig:sets}(c) show a horizontal section of the cone for a fixed $r_0$ and Fig.
\ref{fig:sets}(d) a vertical section. We notice that a necessary condition for a state to
be tripartite entangled is $r_0>2/3$. Below this value, biseparable states fill the cone
completely.

The above criteria can be extended to all states using the \emph{twirling map} that
projects each state onto its rotational invariant subspace:
$ \Pi \rho=\int_G \dd \mathcal {U}\; \mathcal {U} \rho\; \mathcal {U}^{\dagger}$
where $G$ consists of local unitaries $\mathcal {U}=U\otimes U\otimes U$. The following
statements now hold: 
(i) $\Pi \rho$ is $SU(2)$ invariant (ii) if $\rho$ is separable then $\Pi \rho$ is
separable~\cite{Vollbrecht2001}, (iii)  if $\Pi\rho$ is biseparable but not separable then $ \rho$ is not
separable and (iv) if $\Pi \rho$ is genuine tripartite entangled so is $\rho$.

The geometrical description of the Hilbert space depicted in Fig.~\ref{fig:sets}
facilitates the construction of a multipartite entanglement witness, i.e., an observable
$W$ such that $\Tr(W\rho)\ge 0$ $\forall \rho\in \mathcal B$,
and there exists at least one state $\rho\in\mathcal T$ such that $\Tr(W\rho)< 0$. It is
sufficient to choose a witness of the form $W = \sum_i c_i R_i$, where $i \in
\{+,0,1,2\}$, so its expectation value with a rotationally invariant state simplifies to a
scalar product $\Tr(W\rho) = \sum_i c_i r_i \equiv \vec{c} \cdotp \vec{r}$.
We determine $\vec{c}$ using the geometric description of the witness as a plane
intersecting the cone of rotationally invariant states and tangent to the set $\mathcal
B$ at the point $\vec P$ with normal vector $\ver{u}$ [see Fig.~\ref{fig:sets}(d)]. We
find $\Tr(W\rho)=\ver{u} \cdotp (\vec{r}-\vec{P})$ and, from the definition of $\vec{c}$,
we obtain $W=u_0 R_0 + u_1 R_1 + u_2 R_2 -\ver{u}\cdotp \vec{P}$.
The witness plane is calculated with $\vec{P}$ at the midpoint of the line between the
biseparable subsets $\mathcal {B}_2$ and $\mathcal {B}_3$. It also clearly depends on the
choice of $r_0$. In fact
\begin{eqnarray}\label{eq:Witness}
W(r_0)&=&\left(1+\frac{\sqrt{3}}{2}\frac{2-3r_0}{\sqrt{-1+4 r_0-3
r_0^2}}\right) R_0\nonumber\\
&-&R_1-\left(\frac{\sqrt{3}(1-2 r_0)}{2\sqrt{-1+4r_0-3
r_0^2}}+\frac{1}{2}\right)\id.
\end{eqnarray}
The witness can then be rotated about the $r_0$ axis by $\pm 2\pi/3$ to obtain witness
planes tangential to $\mathcal B$ on the lines between $\mathcal {B}_1$ and $\mathcal
{B}_2$, or $\mathcal{B}_1$ and $\mathcal {B}_3$, respectively. The explicit derivation of
Eq.~\eqref{eq:Witness} and the demonstration that $W$ is a witness can be found in
the Appendix \ref{app:witness}.

\section{Entanglement in the spin-$1/2$ XXZ model}\label{sec:XXZ}
Let us now focus  on the study of GME in an open chain of size $N$ described by the XXZ Hamiltonian
\begin{equation}
 H_{\rm XXZ}= \sum_{i=1}^{N-1}\left[J(\sigma_i^{x}
\sigma_{i+1}^{x}+\sigma_i^{y}
\sigma_{i+1}^{y})+\lambda \sigma_i^{z} \sigma_{i+1}^{z}\right],
\label{XXZ}
\end{equation}
where $\lambda$ is the anisotropic parameter. For lattices with even $N$  (and also in the
thermodynamic limit) the sign of $J$ is irrelevant since it is possible to change 
$\sigma_i^{x}\rightarrow -\sigma_i^{x}$ and
$\sigma_i^{y}\rightarrow -\sigma_i^{y}$ for all even (odd) sites, thus the Hamiltonian is
$SU(2)$ invariant for $\lambda/J=\pm 1$. For any value of $\lambda$, $H_{\rm XXZ}$ is
$U(1)$ invariant. For non-bipartite lattices (or odd $N$) geometrical frustration can
appear. In what follows, we assume without loss of generality that $J=1$.

\begin{figure}[t!]
\centering
\qquad\includegraphics[width=\columnwidth]{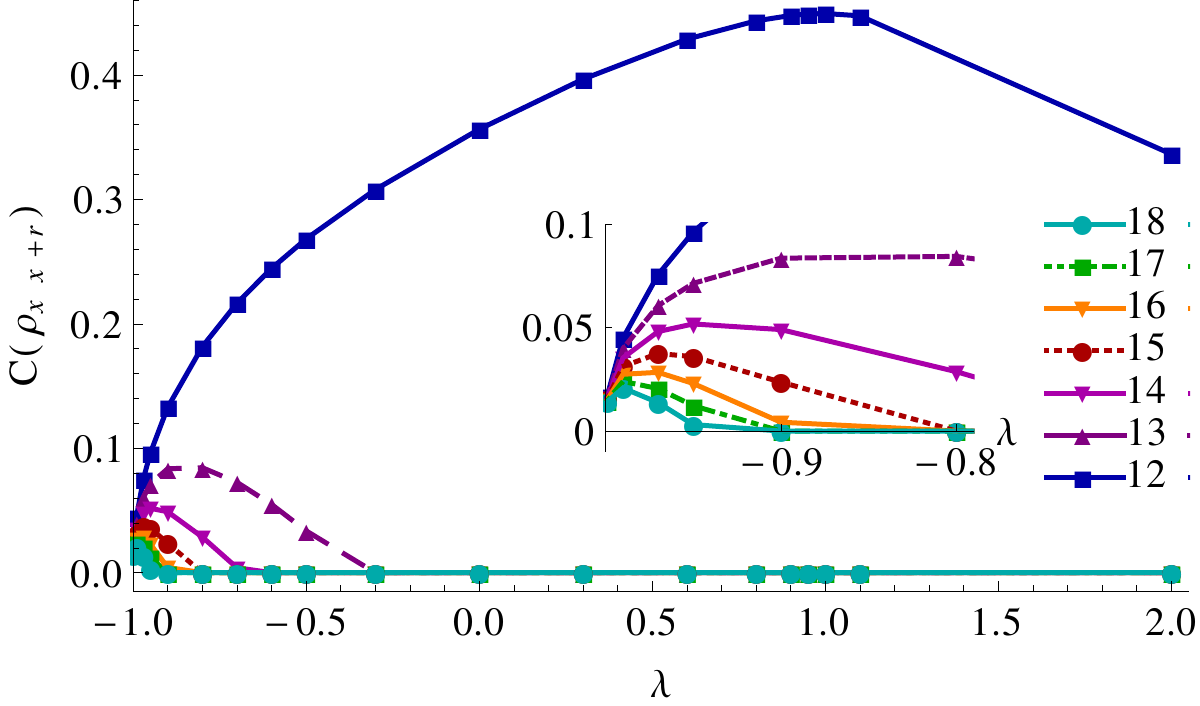}\\
\caption{(Color online) Concurrences of a reduced state of two non-adjacent spins $(x, x+r)$. For $N\gg 1$ all the concurrences collapse to the same value: $1/(N-1)$ close to the isotropic ferromagnetic point ($\lambda=-1^{+}$). The reduced density matrices have been calculated using DMRG simulations with $N=192$. For convenience the spin arrangements are denoted as $12, 13, \ldots$ (corresponding to a pair of nearest-neighbor spins, next to nearest-neighbor spins, etc.), however the entanglement is studied close to the middle of the chain. In the chain of length N the spins correspond to $x\,x+1, x\,x+2,\ldots $ with $x=N/2-3$. The plotted quantity is dimensionless.} 
\label{fig:conc}
\end{figure}
The complete phase diagram of the model is well known: the ground state is ferromagnetic
for $\lambda\le-1$, XY-critical for $-1<\lambda<1$, and Ising-antiferromagnetic (N\'eel)
otherwise. At $\lambda=-1$ a first order transition separates the ferromagnetic and
critical phases. This point is not conformal and has recently attracted some attention
\cite{verrucchi,lauchli2012}. Here we focus on the multipartite entanglement content in
the vicinity of this phase transition. Before proceeding further, notice that at  
$\lambda=-1$ the highly degenerate ground state is in the $SU(2)$ isotropic
ferromagnetic multiplet spanned by any state with maximum total angular momentum
$J$: $\ket{J=N/2,J_{z}}$ for all possible values of $J_{z}$. 
Neither of these states exhibit finite size corrections to the energy per site, nor are
they rotationally invariant. However, each of them corresponds to a
symmetric Dicke state, i.e.,
\begin{equation}
\ket{J=\frac{N}{2},J_{z}=\frac{2k-N}{2}}=\frac{1}{\sqrt{C^{N}_{k}}}\sum_{\mathcal{P}}
\ket{\mathcal{P}({1}^{(k)},0^{(N-k)})},
\end{equation}
with $k$ subsytems in the state $\ket{1}$ and the remaining in the state $\ket{0}$; $\mathcal{P} $
are the elements of the permutation group, $C^{N}_{k}$ is the binomial
coefficient, and $\{\ket{1},\ket{0}\}$ is the computational basis (spin up, spin down).
Thus, the interchange between any two spins leaves the corresponding Dicke state unchanged
and by construction any linear combination as well. In the region $-1<\lambda<1$, the
total spin,  $J$, is not well defined but $J_{z}=0$. In the limit $\lambda\rightarrow
-1^{+}$ the ground state was found numerically to be an equally weighted superposition of
all the elements of the standard basis within the sector  $J_{z}=0$ \cite{verrucchi}.
Finally, notice that for $\lambda=+1$, the ground state is a rotationally invariant
singlet with $J=0$. 

\begin{figure}[t!]
\centering
(a)
\includegraphics[width=\columnwidth]{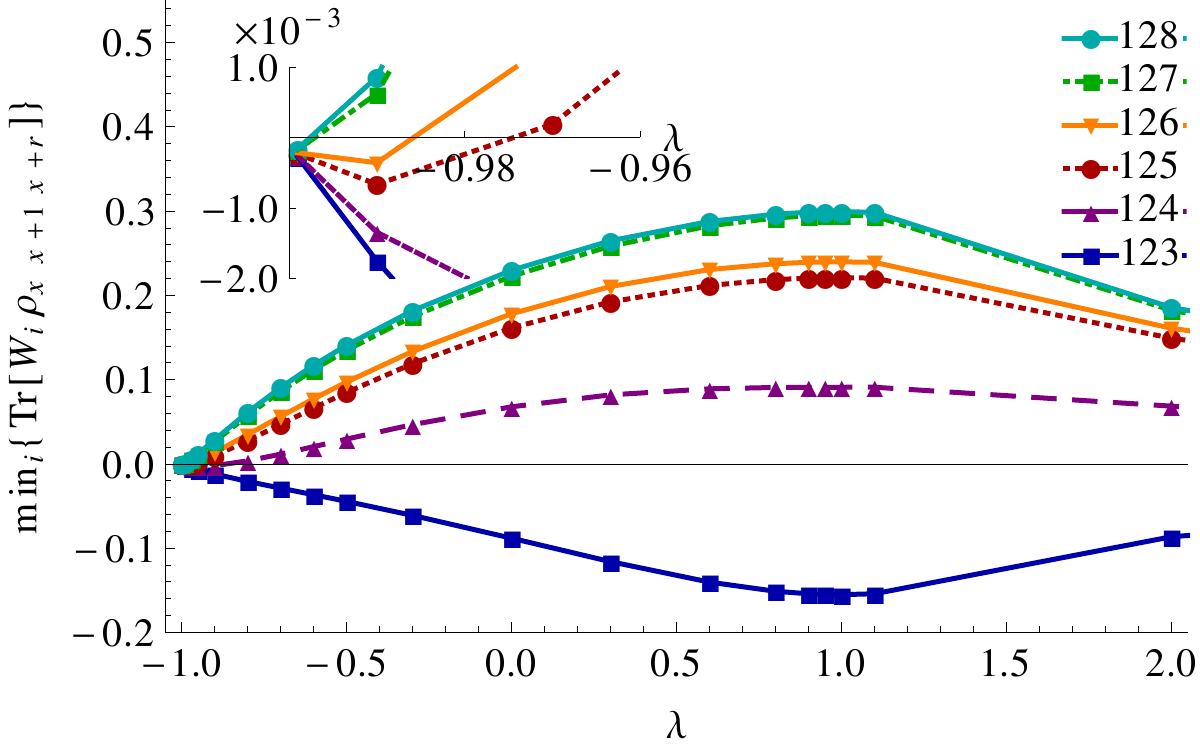}

(b)
\includegraphics[width=\columnwidth]{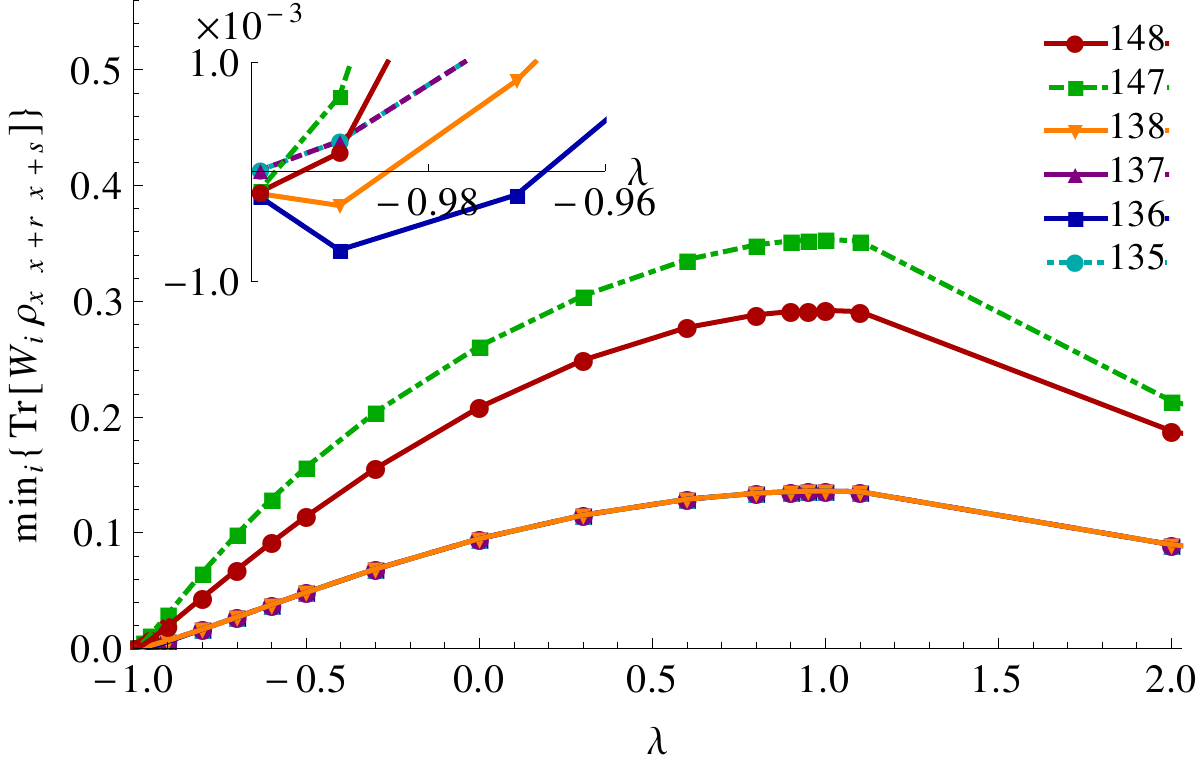}
\caption{(Color online) Mean value of the rotationally-invariant entanglement witness detecting genuine tripartite entanglement for reduced states of three non-adjacent spins (a) $(x, x+1, x+r)$ (b) $(x,x+r,x+s)$. As $\lambda\to-1^{+}$ (see the insets) GME is detected in all the arrangements depicted in (a), and those shown in (b) except for $(135),(137)$. The reduced density matrices have been calculated using DMRG simulations with $N=192$. For convenience the spin arrangements are denoted as $123, 135, \ldots$, however the entanglement is studied close to the middle of the chain. The plotted quantities are dimensionless.} 
\label{fig:witness}
\end{figure}
We now analyze the entanglement content of the model by computing both the bipartite
concurrence $C_{N}(\rho_{i\,i+r})$ of two qubits at distance $r$ as well as
the mean value of the entanglement witness (\ref{eq:Witness}), $\Tr(W\rho_{i\,j\,k})$, for three qubits $(i,j,k)$. Concurrences can be analytically obtained in the $SU(2)$ multiplet after realizing that
any reduction of a multipartite symmetric Dicke state $\ket{N/2,k}$ is also symmetric,
i.e. independent of $r$ and $i$, and read~\cite{Wang_Molmer}:
\begin{eqnarray}
& &C_{N}(\rho_{i\,i+r})=2\rm{max}(0,\rho_{01,01}-\sqrt{\rho_{00,00}\rho_{11,11}})\\
& &=\frac{(N^{2}-4J_z^{2})-\sqrt{(N^{2}-4J_z^{2})[(N-2)^{2}-4J_z^{2}]}}{2N(N-1)} \nonumber
\end{eqnarray}
since the only non-zero matrix elements $\rho_{i\,i+r}$ are the symmetric ones given by
$\rho_{00,00}(\rho_{11,11})=(N \pm 2k)(N-2\pm 2k)/(4N(N-1))$,
$\rho_{{01,01}(01,10)(10,01)(10,10)}=(N^{2}-4k^{2})/ (4N(N-1))$.
As previously noted, the value of the concurrence for large $N$ is very small and close
to the exact value $C_{N}=1/(N-1)$ achieved for  $J_z=0$ ($k=N/2$).
Thus, for large $N$, all members of the $SU(2)$ multiplet (except the trivially separable
$\ket{N/2,\pm N/2}$) have equal concurrences which tend 
to zero in the thermodynamic limit.

Let  us now discuss our results. We compute the ground state of $H_{\rm XXZ}$ for the whole
phase diagram using the density matrix renormalization group (DMRG)~\cite{dmrg} for open
chains of up to 192 sites (We have checked that the accuracy of the results does not depend on the size of the chain). From the ground state, we construct the reduced density
matrix of either two or three spins (not necessarily adjacent) close to the centre of the
chain to avoid edge effects. We then calculate the corresponding concurrences as well as the tripartite entanglement by means of the entanglement witness (\ref{eq:Witness}) for the whole phase diagram.
There exist suitable multipartite entanglement witnesses that detect GME of adjacent sites for two-body models~\cite{Guhne2005,Guhne2006} or global GME of Dicke states~\cite{krammer09,Bergmann13}. Our method, however, allows us to
choose our reduced system at will without imposing further symmetries. Our results are summarized in Fig.~\ref{fig:conc}, where we display $C(\rho_{i\,i+r})$ for
different values of $r$ as a function of $\lambda$, particularly near the isotropic point $\lambda=-1$. In accordance with previous results based on bipartite measures \cite{verrucchi}, we observe that close to the isotropic ferromagnetic point (i.e. $\lambda=-0.999$) all concurrences  for $N\gg 1$ collapse to the same value $1/(N-1)$.
To investigate whether or not this feature of bipartite entanglement is also shared  by the multipartite structure, we investigated the minimum mean value of the witness $\Tr(W\rho_{i\,i+r\,i+s})$ for different spin arrangements (the spin $i$ is near the middle of the chain to avoid edge effects). For adjacent sites $(i,i+1,i+2)$ we recover previous results~\cite{Guhne2005}, indicating the presence of GME for $\lambda>-1$ (see Fig. \ref{fig:witness}(a)). Long-range GME is also detected by the witness for other spin-arrangements $(i,j,k)$ when at least two site indexes have different parity (see Fig. \ref{fig:witness}(b)).
Our results provide evidence that (i) distant multipartite entanglement is present in the system; (ii) the global $SU(2)$ symmetry is already broken very close to the QPT, evidencing the sensitivity of GME to the fine ground state structure.

\section{Summary}\label{sec:summary}
As we demonstrated for the ground state of the XXZ model, the method we provide is not
restricted to $SU(2)$ invariant states and can be applied to any 3-qubit states.
It can also be extended to ground and thermal states of spin-1/2 Hamiltonians in any
lattice geometry, not only 1D. Notice however that while the detection of GME for $SU(2)$
invariant states is unambiguous, the negativity of the expectation value of the witness
$\av{W}{}$ is in general only a sufficient condition for GME.

We have constructed a witness for genuine tripartite entanglement that, although explicitly designed for rotationally invariant states, is useful to assess the
multipartite quantum correlations of states lying outside such a class. Our method is general and can be used together with analytical or numerical methods, e.g. DMRG, exact diagonalization, quantum Monte Carlo or, more generally, any technique yielding three point correlation functions. We have used such
a tool to gather insight into the structure of quantum correlations of the many-body ground state of a spin chain close to a first order quantum
phase transition, beyond the standard framework of bipartite entanglement (see also \cite{Hofmann13}).

\begin{acknowledgements}
We are indebted to O. G\"uhne and M. Hofmann for clarifying insights. We thank R. Augusiak, J. Calsamiglia, A. L\"auchli, R. Mu\~noz-Tapia, and J. Serra for useful discussions. We acknowledge financial support from the Spanish MINECO (FIS2008-01236), European Regional development Fund, Generalitat de Catalunya Grant No. SGR2009-00347. JS acknowledges the support of Fundaci\'o Catalunya -- La Pedrera and Spanish MICINN (FIS208-00784 TOQATA). MP acknowledges the UK EPSRC for a Career Acceleration Fellowship and a grant from the "New Directions for EPSRC Research Leaders" initiative (Grant No. EP/G004759/1). GDC and MP thank the EPSRC funding under the scheme: EP/K029371/1 and the John Templeton Foundation (grant ID 43467) for financial support.
\end{acknowledgements}

\appendix
\section{Construction of the witness operator}
\label{app:witness}
A plane tangent to the surface of the set of biseparable states is uniquely characterised
by a
point $\vec{P}$ and a normal unit vector $\ver{u}$. We choose the point lying on the line
between sets $\mathcal{B}_2$ and $\mathcal{B}_3$ (later denoted by $\ell$) and define a
curve
parametrized by $r_0$ as
\begin{eqnarray}
 \vec{P}(r_0)\!=\!\!\left[r_0,\frac{-1+2 r_0+\sqrt{3} \sqrt{-1+4 r_0-3
r_0^2}}{2},0\right]
\end{eqnarray}
with $r_0>2/3$.
Then the normal vector $\ver{u}$ is the cross-product of the vector tangent to the curve
$\vec{P}(r_0)$:
\begin{equation}
 \vec{v}_1=\frac{\dd}{\dd r_0} \vec{P}(r_0)=\left[1,1+\frac{\sqrt{3}}{2}\frac{2-3
r_0}{\sqrt{-1+4 r_0-3 r_0^2}},0\right],
\end{equation}
and the vector $\vec{v}_2=(0,0,1)$ parallel to the $\ell$.
Therefore, we have
\begin{eqnarray}\label{eq:normal}
 \ver{u}&=&\frac{\vec{v}_1 \times \vec{v}_2}{\|\vec{v}_1\|}\nonumber\\
 &=&\frac{1}{\|\vec{v}_1\|}\left[1+\frac{\sqrt{3}}{2}\frac{2-3r_0}{\sqrt{-1+4 r_0-3
r_0^2}},-1,0\right].
\end{eqnarray}
Substituting Eq. (\ref{eq:normal}) in the expression of the witness:
\begin{equation}
W =u_0 R_0 + u_1 R_1 + u_2 R_2 -\ver{u}\cdotp \vec{P},
\end{equation}
 we obtain the unnormalized witness in the following form
\begin{eqnarray}\label{eq:Wfull}
W(r_0)&=&\left(1+\frac{\sqrt{3}}{2}\frac{2-3r_0}{\sqrt{-1+4 r_0-3
r_0^2}}\right) R_0\nonumber\\
&-&R_1-\left(\frac{\sqrt{3}(1-2 r_0)}{2\sqrt{-1+4r_0-3
r_0^2}}+\frac{1}{2}\right)\id.
\end{eqnarray}
The other two witnesses obtained by rotation of the witness plane by $\pm 2\pi/3$ are
\begin{eqnarray*}\label{eq:Wfull1}
W(r_0)&=&\left(1+\frac{\sqrt{3}}{2}\frac{2-3r_0}{\sqrt{-1+4 r_0-3
r_0^2}}\right) R_0
\nonumber\\
&-&\left(\frac{1}{2}R_1\pm\frac{\sqrt{3}}{2}R_2\right)
\nonumber\\
&-&\left(\frac{\sqrt{3}(1-2 r_0)}{2\sqrt{-1+4r_0-3
r_0^2}}+\frac{1}{2}\right)\id.
\end{eqnarray*}

By construction the operator $W$ is positive for all biseparable rotationally invariant
states, even those with an imaginary component since $\Tr (W R_3)=0$. Furthermore,
$W$ is an entanglement witness for genuine tripartite entangled states, i.e., it is
positive for all biseparable states. For this purpose it is enough to show that 
$\bra{e f}W\ket{e f}\geq 0$ for all $\ket{ef}$ being product vectors with respect to
bipartitions $12|3$, $1|23$, $13|2$. Without loss of generality we focus on the specific
partition and have:
\begin{eqnarray}\label{eq:wit_twirl}
&\, &\bra{e_{12} f_3}W\ket{e_{12} f_3}=\bra{e_{12} f_3}\int \dd {\mathcal U}\, {\mathcal
U}^{\dagger} W {\mathcal U}
\ket{e_{12} f_3}\nonumber\\
&\,&=\Tr (\Pi \ketbra{e_{12} f_3}{e_{12} f_3} W)\geq 0
\end{eqnarray}
where we use the fact that $W$ is rotationally invariant and that the state $\Pi
\ketbra{e_{12} f_3}{e_{12} f_3}$ is rotationally invariant and either biseparable or
separable. Note that Eq. (\ref{eq:wit_twirl}) establishes equivalence between the
biseparability test based on the witness and the twirling criterion for all real density
matrices.


\end{document}